\documentclass[a4paper]{article}
\usepackage{spconf,amsmath,graphicx}
\usepackage{amsfonts}

\usepackage{amssymb}
\usepackage{url}
\usepackage{cite}
\usepackage{hyperref}
\usepackage{booktabs}

\newcommand{\tabincell}[2]{\begin{tabular}{@{}#1@{}}#2\end{tabular}}

\title{MODELING MULTI-SPEAKER LATENT SPACE TO IMPROVE NEURAL TTS:\\
        QUICK ENROLLING NEW SPEAKER AND ENHANCING PREMIUM VOICE}
\name{ Yan Deng, Lei He, Frank Soong}

\address{Microsoft, China\\{\small \tt \{yaden, helei, frankkps\}@microsoft.com}}

\begin{document}
%
\maketitle
\begin{abstract}
Neural TTS has shown it can generate high quality synthesized speech. In this paper, we investigate the multi-speaker latent space to improve neural TTS for adapting the system to new speakers with only several minutes of speech or enhancing a premium voice by utilizing the data from other speakers for richer contextual coverage and better generalization. A multi-speaker neural TTS model is built with the embedded speaker information in both spectral and speaker latent space. The experimental results show that, with less than 5 minutes of training data from a new speaker, the new model can achieve an MOS score of 4.16 in naturalness and 4.64 in speaker similarity close to human recordings (4.74). For a well-trained premium voice, we can achieve an MOS score of 4.5 for out-of-domain texts, which is comparable to an MOS of 4.58 for professional recordings, and significantly outperforms single speaker result of 4.28.

\end{abstract}
\section{INTRODUCTION}
\label{sec:intro}

In the past few years, there have been significant research progresses in neural TTS modeling with an end-to-end structure, e.g. Tacotron+WaveNet \cite{Oord2016, Wang2017, Shen2018}, Char2Wav \cite{Sotelo2017}, DeepVoice \cite{Arik2017} and VoiceLoop \cite{Taigman2018}. These models are all trained directly from text-speech data pairs, via sequence-to-sequence mapping with an attention mechanism. This approach can bypass the need of a well-designed linguistic features front-end and the acoustic feature used in the traditional HMM or DNN/LSTM based system. It is also much easier to expand the system to different languages and speakers. Recent research has proven that neural TTS can generate natural speech with high fidelity that are close to natural recordings for in-domain test sentences \cite{Shen2018}.

Although neural TTS can generate state-of-the-art natural sounding speech, it needs a much larger amount of training data to train a stable and high-quality model than the traditional HMM/DNN/LSTM approach. Usually, a corpus of around 15 hours of speech may still not be enough to train a good end-to-end model. Moreover, there is a key challenge for end-to-end neural TTS is its generalization ability. Degradation of naturalness on synthesizing an out-of-domain sentence does happen, particularly for a long sentence with a rather complex context. For all these problems, adding more training data is a brute force solution. But such heavy data requirements can't be satisfied by using a single speaker corpus for which the data is always limited. We can consider augment the training data by combining corpus of multiple speakers to train a multi-speaker model, which has proven to be a good way to relief the dependency on data size for end-to-end neural TTS modeling \cite{Arik2017NIPS}. Another benefit of multi-speaker modeling is to create customized voice for different speakers using small corpus via speaker adaptation, which has been widely used in traditional TTS systems \cite{Yamagishi2007, Fan2015, Li2016}. But the naturalness and speaker similarity are all far from perfect. And the adapted voice also sounds a bit muffle due to vocoder effect. If we can create high quality customized voice with low resource settings, it will be very attractive for some TTS related applications such as role playing in fairytale and speech-to-speech translation. However, more efficient modeling technology is needed for high quality customized voice creation. Multi-speaker neural TTS will be promising for such problem.

There are already some researches on multi-speaker neural TTS modeling \cite{Taigman2018, Arik2017NIPS, Ping2018, Arik2018, Nachmani2018, Jia2018}. For all these systems, they rely on speaker embeddings to support training multi-speaker model using multi-speaker corpus. In DeepVoice2 and DeepVoice3, they focus on building a multi-speaker model which can generate speech from different voices with less data than single-speaker model \cite{Arik2017NIPS, Ping2018}.But the MOS of naturalness is only about 3.7 for seen speakers in training set, this is a bit low compared with results in \cite{Jia2018}. And the results are similar for VoiceLoop \cite{Taigman2018, Nachmani2018}, still not good in terms of naturalness. Then, they try few-shot speaker adaptation, which can generate speech of new speakers using only a few seconds of speech \cite{Arik2018, Nachmani2018, Jia2018}. Among them, the best results can be obtained in \cite{Jia2018} and the average MOS score is above 4.0 in naturalness for both seen and unseen speakers. But the speaker similarity is still not good enough, especially for unseen speakers, which blocks the real applications.

In this paper, we explore the multi-speaker neural TTS modeling in two different aspects that has not been fully investigated by previous research. The first one is to adapt to new speakers using only several minutes of speech from each speaker, with the goal of creating high quality voice with high similarity. This has been partly touched in \cite{Arik2018, Nachmani2018}. We will continue our work based on their experiments, try to improve the quality and get close to human recording similarity. The second one is to enrich a premium voice for context reinforcement, with the goal of getting close to human quality voice with better generalization to out-of-domain inputs, especially for long sentences with complex context. To the best of our knowledge, there is no such research on how to enrich a premium voice using multi speaker end-to-end neural TTS modeling and we are the first to investigate on it.

The rest of the paper is organized as follows. Section \ref{sec:system} describes the framework of our multi-speaker neural TTS system, including the details of speaker adaptation. Section \ref{sec:exper} shows the TTS experiments and subjective evaluation results. We give a brief conclusion in Section \ref{sec:conc}.

\section{MULTI-SPEAKER NEURAL TTS}
\label{sec:system}

\begin{figure}[t]
\includegraphics[width=8.5cm,height=6cm]{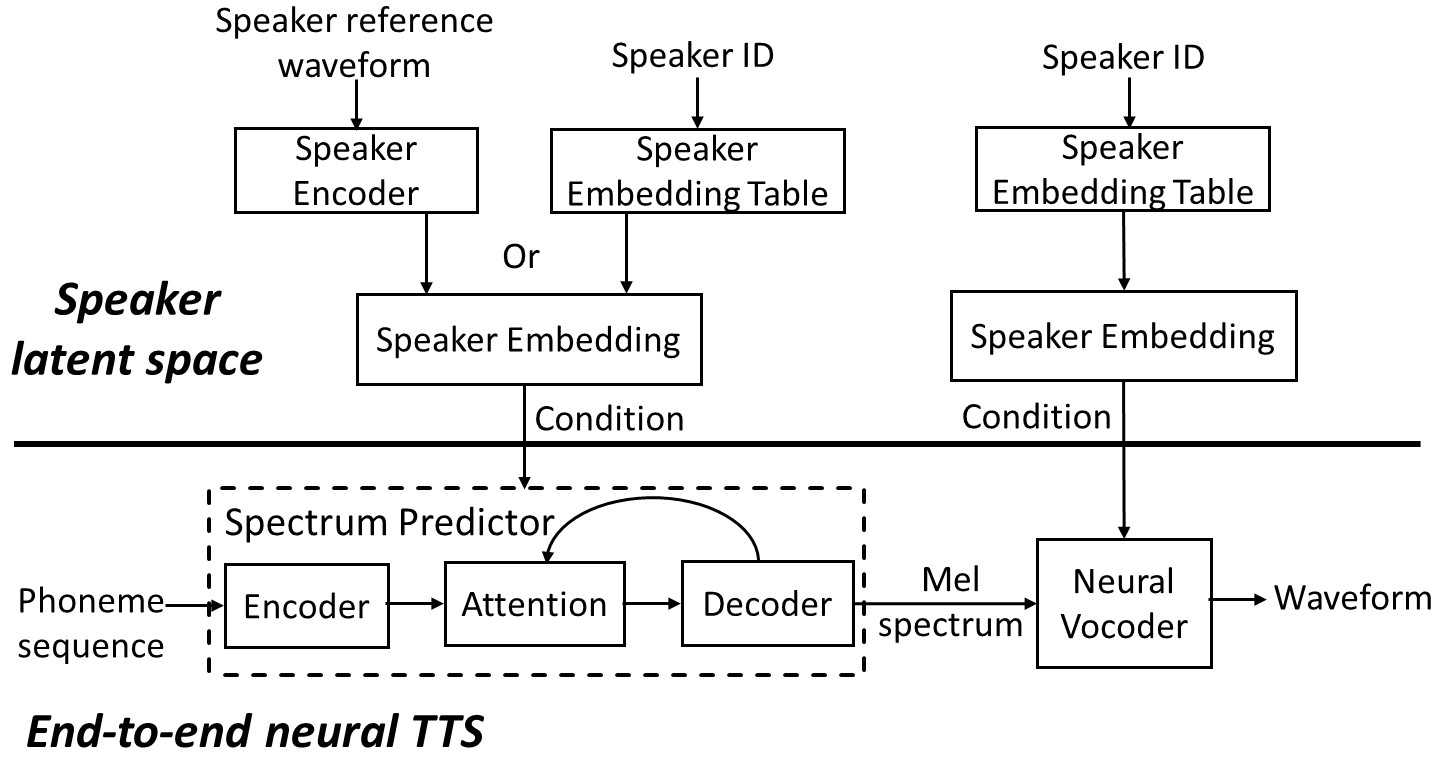}
\caption{\label{fig:FIG1}Proposed multi-speaker neural TTS system.}
\end{figure}

Speaker embedding is very effective in modeling the discriminative information in speaker latent space \cite{Li2017}. It has been used in multi-speaker modeling for TTS to control the speaker characteristics in generated speech \cite{Taigman2018, Arik2017NIPS, Ping2018, Arik2018, Nachmani2018, Jia2018}. We will adopt speaker embedding to model the speaker latent space. The proposed multi-speaker neural TTS system includes two main components as illustrated in  Fig.\ref{fig:FIG1}. One is used to extract speaker embedding from the speaker latent space and the other is end-to-end neural TTS. We will describe each component in detail.

\subsection{Modeling speaker latent space}
\label{ssec:speaker}

We use a speaker model to estimate the embedded speaker features from the latent space. In the multi-speaker neural TTS system, the output of the speaker model is used as an additional information to control the speaker characteristics in both training and inference. There are different networks proposed for modeling speakers \cite{Oord2016, Arik2018, Nachmani2018, Jia2018}. We try two variants of speaker models: a fixed lookup-table of speaker embedding or a speaker encoder. The speaker modeling network is jointly trained with spectrum predictor or neural vocoder.

The network in speaker encoder is designed similar with \cite{Nachmani2018, Wang2018}. Mel-spectrum is used as input and the corresponding network consists of a stack of convolution layers, followed by average pooling, several layers of fully connected layers and an affine projection. The speaker embedding is created by L2-normalization of projection output. A speaker encoder can be trained on a large corpus of many speakers, which is for text-independent speaker recognition task. The encoder can thus provide a better estimation of speaker embedding features independent of contents and recording conditions.

\subsection{End-to-end neural TTS}
\label{ssec:neuraltts}

There are two networks in end-to-end neural TTS: spectrum predictor and neural vocoder. They are separately trained and work sequentially in inference.

\subsubsection{Spectrum predictor}
\label{sssec:spectrum}

The spectrum predictor follows the sequence-to-sequence modeling with an attention architecture \cite{Shen2018, Jia2018}. It predicts the corresponding mel-spectrum directly from a text input. We use phoneme sequence as input to reduce pronunciation errors. The embedded speaker information is concatenated to the encoder output and then passed to the attention layer. We use mean square error of predicted spectrum as a loss function. Also, L\textsubscript{2}-regularization is applied with a small weight. For the spectrum predictor, we try both speaker encoder and embedding lookup-table to get the best speaker embedding estimation method.

\subsubsection{Neural vocoder}
\label{ssec:vocoder}

We use the auto-regressive generative model as the vocoder to generate high quality speech output \cite{Oord2016, Jia2018}. It uses the predicted mel-spectrum as conditional information and generates speech sample by sample. The framework is close to the one reported in \cite{Jia2018}, but 20 dilated convolution layers are used instead. Also, different from all previous work, the neural vocoder is conditioned upon the embedded speaker features. We believe the embedded speaker features are critical for generating a highly natural output speech which sounds similar to the target speaker. For neural vocoder, we only optimize a fixed lookup-table to get speaker embeddings for all training speakers and new speakers.

\subsection{Speaker adaptation}
\label{ssec:adapt}

Speaker adaptation is to adapt a pre-trained, multi-speaker model towards a new speaker. In our multi-speaker neural TTS model, there is a separate speaker model for embedded speaker feature extraction. We have two options in speaker adaptation: only updating the speaker model; or updating the entire model set. By comparing the performance of the two approaches, we found that the model is easier to converge when the whole model set is updated, with not less than 50 utterances. Similar results are also reported \cite{Arik2018}. Hence, the entire set of model parameters is updated for both the spectrum predictors and neural vocoders in our experiments.

\section{EXPERIMENTS}
\label{sec:exper}

\subsection{Experimental setup}
\label{sec:confi}

We use internal corpus with high-quality recordings to train multi-speaker neural TTS model. The multi-speaker corpus is composed of English speakers with different accents, including American, British, Australian, Canadian and even Indian accent, about 100 hours of speech in total. For this multi-speaker corpus, we also consider gender, age and content coverage. All audios are down sampled to 16kHz. The beginning and ending silence are all trimmed to a fixed length.

In our experiment, the spectrum predictor and neural vocoder are separately trained with the multi-speaker corpus. We have two different speaker models, one for spectrum predictor and one for neural vocoder. All speaker models are jointly trained with either spectrum predictor or neural vocoder. The dimension of speaker embedding is set to 128 for all models.

We use phoneme input for the spectrum predictor. The network is trained using a batch size of 32 on 4 GPUs. We use Adam optimizer and the initial learning rate is set to 10\textsuperscript{-3}. It starts to decay exponentially after 50000 steps. The minimum learning rate is set to 10\textsuperscript{-5}. For the neural vocoder, we use ground truth mel-spectrum as input and train it on 8 GPUs, using the exponential moving average strategy to update network parameters. The learning rate is kept constant to 10\textsuperscript{-4}. We use the mixture of logistic distribution to model 16-bit samples at 16kHz.

Subjective listening tests are used for evaluating the TTS quality of all experiments. The 5-point Mean Opinion Score (MOS) and AB preference test are used to evaluate the naturalness. For adaptation to new speakers, we also use 5-point MOS score to evaluate the speaker similarity.

\subsection{Enrolling new speakers with small dataset}

Our first experiment is to adapt to new speaker using only a few minutes of the target speaker's speech. Here, speaker encoder composed of convolutional layer and fully-connected layer, is used in spectrum predictor to estimate speaker embedding based on our preliminary results. We use 6 convolutional layers with 3x3 kernel, the channel in each layer is (32, 32, 64, 64, 128, 128). Then average pooling along time is applied, followed by two fully-connected layers of 256 in size. Finally, an affine projection and L\textsubscript{2}-normalizaiton is used to get the speaker embedding vector. Here, we set the dimension of embedding vector to 128. In this experiment, we select eight non-professional English speakers as target speaker. Each speaker has only about 50 utterances (3$\sim$4 minutes of speech).

We do MOS test for both naturalness and speaker similarity to evaluate the performance of neural TTS. In this test, we use 40 utterances for each system to evaluate naturalness and 10 utterances for each to evaluate speaker similarity. Every utterance is listened by at least 15 judges. Results are shown in Table. \ref{tab:mos1}. These are average MOS scores on target speakers. For comparison, we also give the MOS of real recordings. The results show that neural TTS can get very satisfied performance, the MOS on naturalness is above 4.1, which is very good for speakers with only 50 utterances to do adaptation. In terms of speaker similarity, we have obtained the best results so far on the speaker adaptation task with only several minutes of speech. The MOS score for speaker similarity is 4.64, very close to that of natural recordings (4.74).

\begin{table}[th]
  \caption{MOS for naturalness and speaker similarity on in-domain test set with 95\% confidence intervals}
  \label{tab:mos1}
  \centering
  \begin{tabular}{ccc}
    \toprule
    \textbf{}      & \textbf{Recording}      & \textbf{Neural TTS} \\
    \midrule
    {Naturalness}       &4.57\underline{+}0.1      &4.16\underline{+}0.07           \\
    {Similarity}       &4.74\underline{+}0.1      &4.64\underline{+}0.07           \\
    \bottomrule
  \end{tabular}
\end{table}

\begin{figure}[t]
\includegraphics[width=8.5cm,height=8.5cm]{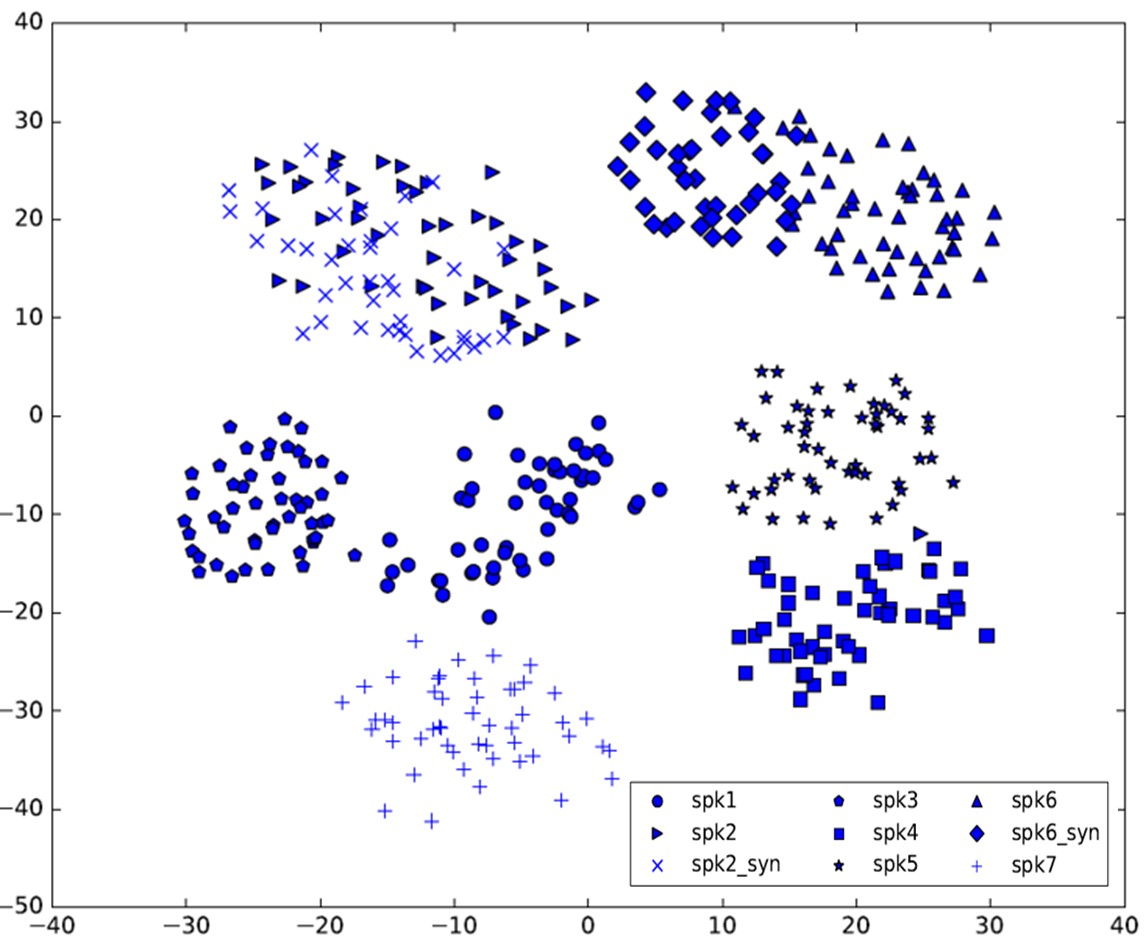}
\caption{\label{fig:FIG2}Visualization of speaker embeddings extracted from real (spk1-7) and synthetic utterances (spk2/6\underline{ }syn).}
\end{figure}

We also give the t-SNE visualization of estimated speaker embeddings in Fig.\ref{fig:FIG2}. It shows that points of different speakers are well clustered and separated from each other. But points from adapted voice lie very close to real recordings of the corresponding speaker (spk2 vs spk2\underline{ }syn and spk6 vs spk6\underline{ }syn). Especially for spk2, it's very hard to separate the two clusters in the space. This means that our speaker encoder has learned a good representation of the speaker space.

\subsection{Enhancing premium voice}

The second experiment is to enrich a premium voice which has about 15 hours of recorded speech from a professional speaker. The target speaker is included in the multi-speaker corpus. We do adaptation on the target speaker's corpus for further refinement. In this experiment, we use the lookup-table to estimate speaker embeddings in spectrum predictor.  Here, the dimension of speaker embedding vector is set to 128. We use the target speaker neural TTS model as the baseline, which is trained only on the target speaker's corpus.

\begin{figure}[t]
\includegraphics[width=8.5cm]{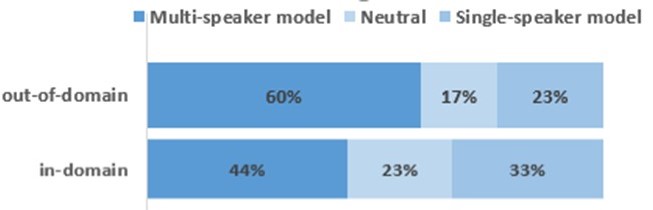}
\caption{\label{fig:FIG3}AB preference test scores for single-speaker neural TTS and multi-speaker neural TTS in different domains.}
\end{figure}

The premium voice without adaptation, has achieved a rather high-quality synthesis for the in-domain texts, with MOS gap to recording 0.09.  In the TTS subjective tests, we use one in-domain set (40 sentences) and one out-of-domain set (100 sentences, including some long sentences with complex context up to 50 words). The results are shown in Fig. \ref{fig:FIG3}. The multi-speaker adapted model is consistently better than the single-speaker one (at \textit{p} $<$ 0.01 level). The improvement gets more significant for out-of-domain texts. This demonstrates that multi-speaker neural TTS model has a better generalization to out-of-domain sentences than the single-speaker model, especially for longer sentences with a more complex context. We also do a MOS evaluation for the multi-speaker neural TTS model on above out-of-domain test sets. The results are shown in Table. \ref{tab:mos3}. Our multi-speaker model can achieve an MOS of 4.5 while single-speaker model can only receive an MOS of 4.28. And the MOS of multi-speaker model is very close to that of recording (4.58). Our proposed multi-speaker model can greatly improve the generalizability to out-of-domain texts.

\begin{table}[th]
  \caption{MOS for naturalness on out-of-domain test set with 95\% confidence intervals.}
  \label{tab:mos3}
  \centering
  \begin{tabular}{ccc}
    \toprule
    \textbf{Recording}      & \tabincell{c}{\textbf{Multi-speaker}\\\textbf{model}}      & \tabincell{c}{\textbf{Single-speaker}\\\textbf{model}} \\
    \midrule
    4.58\underline{+}0.14      &4.5\underline{+}0.08      &4.28\underline{+}0.11          \\
    \bottomrule
  \end{tabular}
\end{table}

\begin{table}[th]
  \caption{Quality along with training data size (in MOS for naturalness on in-domain test set).}
  \label{tab:mos4}
  \centering
  \begin{tabular}{cccc}
    \toprule
    \textbf{Recording}      & \textbf{Data size}      & \tabincell{c}{\textbf{Multi-speaker}\\\textbf{model}}      & \tabincell{c}{\textbf{Gap to}\\\textbf{recording}} \\
    \midrule
     4.63\underline{+}0.1      & 0.5-hour       & 4.07\underline{+}0.08       & 0.56            \\
                                             & 2-hour       & 4.22\underline{+}0.06       & 0.41             \\
                                             & 5-hour       & 4.34\underline{+}0.06       & 0.29             \\
                                             & 10-hour       & 4.47\underline{+}0.05       & 0.16             \\
                                             & 15-hour       & 4.57\underline{+}0.05      & 0.06             \\
    \bottomrule
  \end{tabular}
\end{table}

Finally, we show the quality curve along with different data size in terms of MOS for naturalness in Table. \ref{tab:mos4}. The quality can achieve an MOS of 4.07 using only 0.5 hours of speech and it keeps improving along with increasing of the training data. With about 15 hours of training data, we can get close to human speech quality using our multi-speaker model.

\section{Conclusion}
\label{sec:conc}

In this paper, we investigate a multi-speaker neural TTS model for improving the speech quality in naturalness and speaker similarity. Our results show that the proposed multi-speaker TTS model with speaker latent space information can make the new neural, speaker adaptive TTS to generate a new speaker's voice with an MOS of 4.16  for its naturalness, and with speaker similarity score very close to natural recordings, using only 50 enrollment utterances.  The new model can also improve the quality of a well-trained, single speaker premium voice trained with about 15 hours of speech. It is versatile enough to generalize to out-of-domain test sentences by exploiting the information encoded in the multi-speaker database in a supervised pretraining way, and achieves an start-of-the-art quality close to professional recordings on out-of-domain test.

\section{Acknowledgement}
\label{sec:ack}

The authors would like to thank Xi Wang, Yanqing Liu and Huaiping Ming for helpful feedback and providing samples of single-speaker neural TTS model in evaluation.

\bibliographystyle{IEEEbib}
\bibliography{strings,refs}

\end{document}